\documentclass[12pt]{article}
\usepackage{multirow}
\usepackage{booktabs}
\usepackage{dsfont}
\usepackage{subcaption}
\usepackage{amssymb}
\usepackage{amsmath}
\usepackage{amstext}
\usepackage{amsthm}
\usepackage{graphics} 
\usepackage{color,verbatim}
\usepackage{mathtools}
\usepackage{enumitem}
\usepackage{algorithm}
\usepackage{algpseudocode}
\usepackage{chngcntr}
\usepackage{xr}
\usepackage[bookmarks=false,
colorlinks,
linkcolor=blue,
anchorcolor=blue,
citecolor=blue,
filecolor=blue,
urlcolor=black
]{hyperref}

\makeatletter
\newcommand*{\addFileDependency}[1]{% argument=file name and extension
\typeout{(#1)}% latexmk will find this if $recorder=0
\@addtofilelist{#1}
\IfFileExists{#1}{}{\typeout{No file #1.}}
}\makeatother

\selectfont

\DeclareMathOperator{\diag}{diag}
\DeclareMathOperator{\sign}{sign}

\DeclareMathOperator{\argmin}{argmin}

\def\squarebox#1{\hbox to #1{\hfill\vbox to #1{\vfill}}}
\def\boxit#1{\vbox{\hrule\hbox{\vrule\kern6pt
			\vbox{\kern6pt#1\kern6pt}\kern6pt\vrule}\hrule}}

\def\boxit#1{\vbox{\hrule\hbox{\vrule\kern6pt
          \vbox{\kern6pt#1\kern6pt}\kern6pt\vrule}\hrule}}

\def\sumk0p{\sum_{k=0}^\ell}

\def\bE{\mathbb{E}}

\renewcommand{\tilde}{\widetilde}
\renewcommand{\hat}{\widehat}

\numberwithin{equation}{section}
\allowdisplaybreaks

\newtheoremstyle{break}% name
  {9pt}%      Space above, empty = `usual value'
  {12pt}%      Space below
  {\upshape}% Body font
  {}%         Indent amount (empty = no indent, \parindent = para indent)
  {\bfseries}% Thm head font
  {.}%        Punctuation after thm head
  {\newline}% Space after thm head: \newline = linebreak
  {}%         Thm head spec

\newtheorem{rem}{Remark}
\newtheorem{theo}{Theorem} 
\newtheorem{prop}{Proposition}

\theoremstyle{definition}
\newtheorem{exa}{Example}
\theoremstyle{break}

\theoremstyle{remark}

\usepackage{natbib}
\title{•}
\author{}
\date{}
\begin{document}
\begin{center}
{\bf \Large 
Supervised Manifold Learning for Functional Data} \\[1cm]
\end{center}

\begin{center}
Ruoxu Tan \\
School of Mathematical Sciences and School of Economics and Management,\\
Tongji University\\
ruoxut@tongji.edu.cn\\ 
Yiming Zang \\
Department of Sciences, \\
North China University of Technology\\
yiming.zang@ncut.edu.cn\\
\end{center}

\begin{abstract}
Classification is a core topic in functional data analysis. A large number of functional classifiers have been proposed in the literature, most of which are based on functional principal component analysis or functional regression. In contrast, we investigate this topic from the perspective of manifold learning. It is assumed that functional data lie on an unknown low-dimensional manifold, and we expect that superior classifiers can be developed based on the manifold structure. To this end, we propose a novel proximity measure that takes the label information into account to learn the low-dimensional representations, also known as the supervised manifold learning outcomes. When the outcomes are coupled with multivariate classifiers, the procedure induces a new family of functional classifiers. In theory, we prove that our functional classifier induced by the $k$-NN classifier is asymptotically optimal. In practice, we show that our method, coupled with several classical multivariate classifiers, achieves highly competitive classification performance compared to existing functional classifiers across both synthetic and real data examples. Supplementary materials are available online.
\end{abstract}

\section{Introduction}
Functional data analysis has become increasingly prominent across diverse scientific fields, including medicine, neuroscience, astronomy, chemometrics, etc. As a result, there is a growing demand for sophisticated models and algorithms designed for functional data analysis. A core topic in functional data analysis is classification, i.e., training a functional classifier based on observed data to predict the label for a future functional observation. There exists a fairly rich literature on this topic. Due to the infinite-dimensional nature of functional data, a large number of existing functional classifiers are built on the functional principal component analysis (FPCA), whose theoretical basis is due to the Karhunen–Lo\`eve expansion: A second order random process $X$ can be decomposed as
\begin{align*}
X(t) = \mu(t) + \sum_{j=1}^\infty \zeta_j\phi_j(t),   
\end{align*}
where $\mu(t) = E\{X(t)\}$, $\zeta_j$ denote the principal component (PC) scores, $\phi_j$ represents the eigenfunctions of the covariance operator, also referred to as the PC basis. A finite number of the PC scores serves as a finite-dimensional approximation of the random process. \citet{Delaigle2013} and \citet{Galeano2015} adapted the linear and quadratic discriminant analysis to the PC scores to develop functional linear and quadratic discriminant classifiers. \citet{Delaigle2012} proposed to project functional data onto a one-dimensional space spanned by a linear combination of the PC basis, and showed that asymptotic perfect classification can be achieved under specific conditions. \citet{Wu2013} treated a finite number of PC scores as predictors and performed the classical support vector machines on the PC scores for functional classification. \citet{Dai2017} proposed to estimate the densities of the PC scores nonparametrically and to utilize the Bayes classifier. \citet{Wang2024} constructed a deep neural network on the PC scores as a functional classifier, and \citet{Wang2023} generalized it to the setting of multidimensional functional data. In addition to FPCA-based methods, other functional classifiers include regression-based methods \citep{Ferraty2003,Leng2006,Preda2007,Araki2009}, distance-based methods \citep{Alonso2012,Kraus2019}, and nearest neighbors \citep{Biau2005,Biau2010}; see \citet{Wang2024a} for a recent review.

In this article, we propose a novel manifold learning procedure for functional data by taking the label information into account, which induces a new family of functional classifiers. Instead of the FPCA or regression, we start from the assumption that functional data lie on an unknown low-dimensional manifold. This assumption for high-dimensional data forms the basis for manifold learning \citep{Meila2024}, and has also been invoked under the functional data setting \citep{Chen2012,Lin2021,Tan2024}. A motivating example of the manifold assumption is that functional data often exhibit phase variation, which inflates the variance of data and distorts the PCs \citep{Marron2015}. To investigate the non-linearity of phase variation, manifold techniques have been developed \citep{Kneip1992,Srivastava2011}. In addition, \citet{Marron2021} is a relatively new monograph that collects many useful techniques to deal with non-Euclidean data, including manifold data analysis, curve registration, and distance-based classification that are relevant to our work.

Based on the manifold assumption, we propose a novel proximity measure that incorporates the label information to quantify distance between functional data. Not only does the proximity measure penalize for different classes, but it also tries to maintain the underlying manifold structure in terms of the geodesic distance. The latter property is crucial to classification, because the manifold learning step being continuous is necessary for an interpolator to accurately predict the low-dimensional representation of a future observation; see Section~\ref{sc_method} for more discussion. Once the proximity measure is obtained for all pair-wise individuals, we perform the multidimensional scaling \citep[MDS,][]{Cox2008} to obtain the low-dimensional manifold learning outcomes. Such outcomes can also be used for data visualization if the target dimension is two or three. Finally, a smooth interpolator accompanied with any reasonable multivariate classifier yields a functional classifier. In theory, we provide the with-in-sample and out-of-sample convergence rates of the estimated coordinate map. As for classification, we also show that our functional classifier coupled with the vanilla $k$-NN classifier is asymptotically optimal. Note that the intrinsic dimension being low is opposite to the number of PCs being high when the manifold is flat. This latter point is a core assumption to prove nice theoretical properties, say, the asymptotic perfect classification, of several functional classifiers \citep{Delaigle2012,Dai2017,Berrendero2018}. Therefore, our investigation of the low intrinsic dimension serves as an interesting complement to the existing literature.

Manifold learning has been an active research field since the seminal works by \citet{Tenenbaum2000} and \citet{Roweis2000}, which has gained tremendous success in bioinformatics \citep{Moon2018}, image processing \citep{Pless2009}, etc. In functional data analysis, \citet{Chen2012} extended the Isomap \citep{Tenenbaum2000} to functional data and developed the functional manifold component; \citet{Tan2024} extended the parallel transport unfolding \citep{Budninskiy2019} to functional data and conducted a cluster analysis. However, their works belong to unsupervised manifold learning, i.e., no label variable is involved, and thus their methods are not suitable for classification. In contrast, our manifold learning procedure is designed to take the label information into account, also known as supervised manifold learning. There is a vast literature on supervised manifold learning for high-dimensional data; see \citet{Chao2019} for a survey. Also, there exist classifiers based on distances penalized by different classes; see e.g., Ch~11.4 of \citet{Marron2021}. Our method may be categorized into Isomap-based methods, but it differs from existing supervised Isomap methods \citep{Vlachos2002,Geng2005,Zhang2018} in two main aspects: Our proximity measure is new, and we are dealing with functional data instead of high-dimensional data.

The rest of this article is organized as follows. In Section~\ref{sc_model}, we first introduce the model and present two examples to show the difficulty of the problem and the motivation of our method. We then introduce our main method in  Section~\ref{sc_method}. In Section~\ref{sc_est}, we discuss a detailed estimation procedure to implement the proposed method, including the selection of tuning parameters. We provide theoretical properties of our estimators in Section~\ref{sc_theo}, and conduct simulation studies and real data applications in Sections~\ref{sc_sim} and \ref{sc_real}, respectively. We conclude in Section~\ref{sc_dis}. 

\section{Functional Supervised Manifold Learning}\label{sc_MM}

\subsection{Model and Synthetic Examples}\label{sc_model}
Consider an independent and identically distributed (i.i.d.) sample $\{(X_i,Y_i)\}_{i=1}^n$ of $(X,Y)$, where $X:\mathcal{T}\to \mathbb{R}$ is a second-order random process on a compact interval $\mathcal{T}$, and $Y\in \mathcal{Y}=\{0\,\ldots,M-1\}$ is the label variable of the $M$ classes. We assume that $X$ lies on an unknown functional manifold $\mathcal{M}$ of intrinsic dimension $d\ll n$, which is regarded as an embedded manifold into $L^2(\mathcal{T})$, the Hilbert space on $\mathcal{T}$. In particular, it is endowed with the metric $d_{\mathcal{M}}$ induced by the $L^2$ metric $\|\cdot\|_{L^2}$ in $L^2(\mathcal{T})$. We aim to construct a classifier that takes advantage of the manifold structure to predict the class of a future observation $X_0$. 
 
It is reasonable to assume that the functional variables from different classes concentrate in different regions of the manifold, i.e., the modes of the distributions $X|Y=y$ are easily separated in $\mathcal{M}$. However, the optimal boundaries for separating classes may be highly nonlinear in the original space, which means that standard linear classification methods may fail. Yet, the optimal boundaries may become simpler if the data are properly embedded in a low-dimensional space. Example~\ref{exa1} illustrates this phenomenon in $\mathbb{R}^3$, and Example~\ref{exa2} is a more concrete example showing the classification difficulty under the functional manifold setting. 

\begin{figure}[t]
\centering
\includegraphics[width=0.4\textwidth]{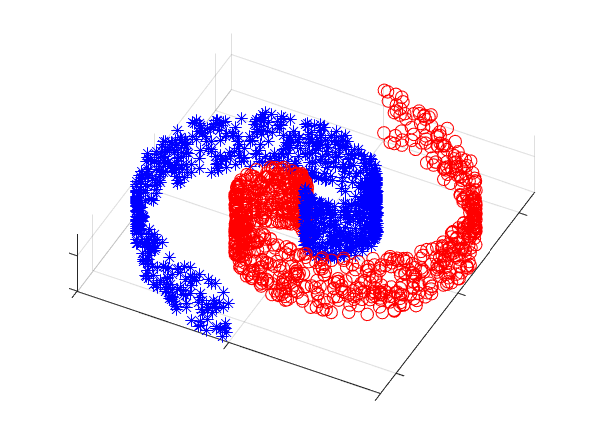}
\includegraphics[width=0.4\textwidth]{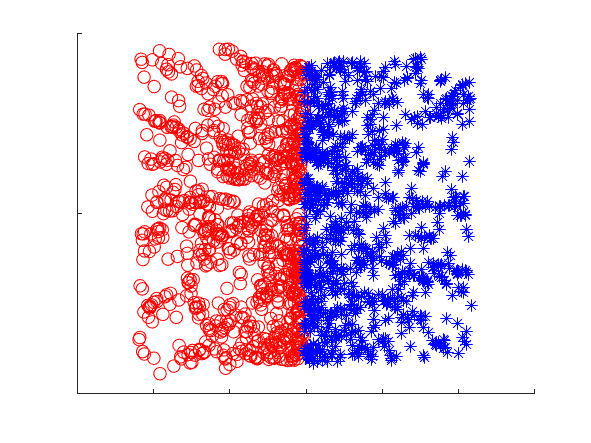}
\caption{Left: the original data of two Swiss rolls ($\circ$ and $*$) representing two classes in $\mathbb{R}^3$; right: the unfolded data using the Isomap in $\mathbb{R}^2$.}\label{fg_exm1}
\end{figure}

\begin{exa}\label{exa1}
The left panel of Figure~\ref{fg_exm1} shows a data cloud in $\mathbb{R}^3$ of two Swiss rolls (circles and stars) representing two classes. There is no plane that can perfectly separate the two Swiss rolls in $\mathbb{R}^3$, but a straight line can do it if the data are unfolded into $\mathbb{R}^2$ using the Isomap \citep{Tenenbaum2000}.
\end{exa}

\begin{rem}
It is interesting to note that Example~\ref{fg_exm1} and the famous motivating example of the kernel trick \citep[see,~e.g.,~Sec.~22.10 of][]{Wasserman2013} show two opposite strategies. In the kernelization example, the nonlinear optimal classification boundary becomes linear by lifting the data to a higher dimensional space, and thus a simple classifier, say the support vector machines (SVM), can be applied on the high-dimensional space. In contrast, in Example~\ref{fg_exm1}, we (nonlinearly) project the data into a lower-dimensional space to find the linear optimal classification boundary. The motivation here is that the data lie on a low-dimensional manifold.
\end{rem}

\begin{figure}[t]
\centering
\includegraphics[width=0.24\textwidth]{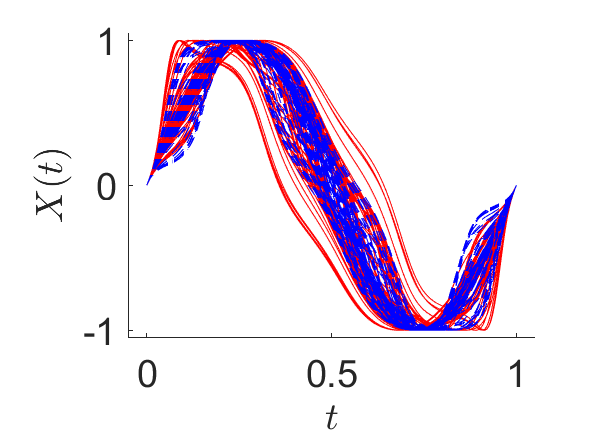}
\includegraphics[width=0.24\textwidth]{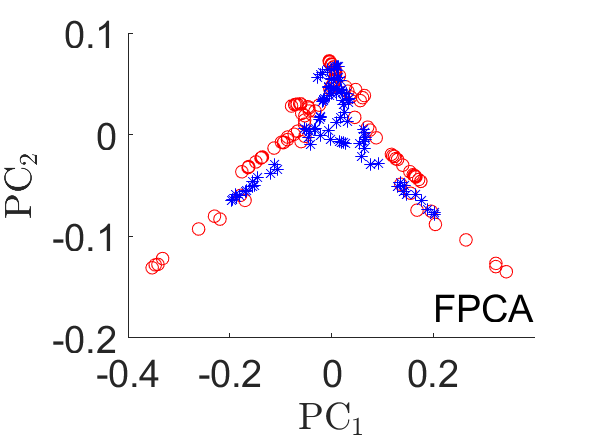}
\includegraphics[width=0.24\textwidth]{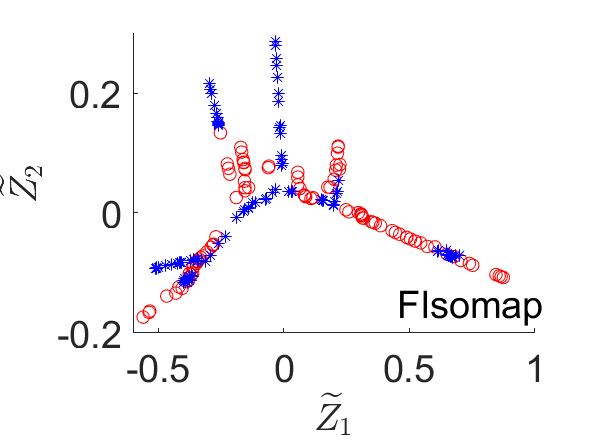}
\includegraphics[width=0.24\textwidth]{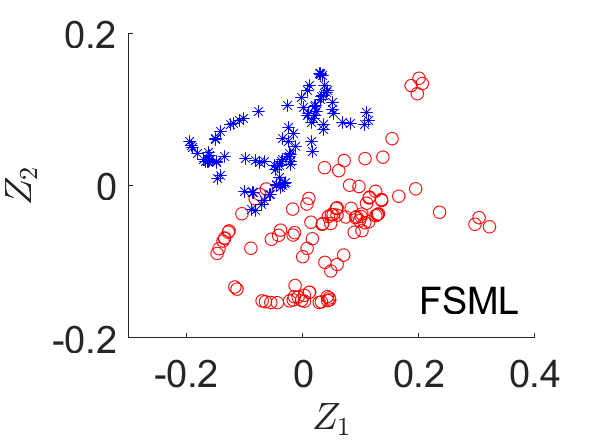}
\caption{The panels from left to right show a simulated sample of curves as in Example~\ref{exa2}, the first two principal component scores (PC$_1$ and PC$_2$) using the FPCA, the unsupervised manifold learning outcomes $\tilde{Z}=(\tilde{Z}_1,\tilde{Z}_2)$ using the FIsomap and the supervised manifold learning outcomes $Z=(Z_1,Z_2)$ using our proposed method FSML, respectively. The class $Y=1$ is denoted by solid line ($-$) and star ($*$), while $Y=2$ is denoted by dashed line ($-$\,$-$) and circle ($\circ$).}\label{fg_exm2}
\end{figure}

\begin{exa}\label{exa2}
Consider a functional time warping model $X(t) = g\{\gamma(t)\}$, for $t\in [0,1]$, which is commonly used for modeling phase variation. We set $\gamma(t) = \int_0^t \exp\{\theta(s)\}\,ds/\int_0^1 \exp\{\theta(s)\}\,ds$, where $\theta(t) = \theta_\omega(t) = \sum_{k=1}^{k_0}  \sin(k \omega ) \sin(\pi kt) / \{(k-8)^2+2\}$ with $\omega \sim \textrm{Unif}[0,2 \pi]$, for $t\in [0,1]$. The functional variable $X$ can be seen as lying on a manifold of dimension one, as it is generated by a one-dimensional variable $\omega$. Yet, the valued space of $X$ is sufficiently complicated: We can show that $\Theta=\{\theta(\cdot)\mid \omega \in [0,2\pi],k_0= \infty\}$, as a subspace of $L^2([0,1])$, cannot be covered by any finite-dimensional linear subspace of $L^2([0,1])$; see Section A of the Appendix for details.  

For the label variable $Y$, we set $Y=1$ if $w\in[0,\pi]$ and $Y=2$ otherwise. The first panel of Figure~\ref{fg_exm2} shows a simulated sample of 200 curves (half in each class) with $g=\sin(2\pi t)$ and $k_0=50$. We apply the FPCA to the sample, whose first two principal component scores are shown in the second panel of Figure~\ref{fg_exm2}. The manifold learning outcomes using the FIsomap \citep{Chen2012} and our proposed
functional supervised manifold learning (FSML) are shown in the third and fourth panels of Figure~\ref{fg_exm2}, respectively.  We see that the PC scores and the FIsomap outcomes from different classes are entangled with each other, while the supervised manifold learning outcomes are more separable by class. Furthermore, although the intrinsic dimension is one, at least five PCs are required to explain 95\% variances of the data, indicating inefficiency of the FPCA. 
\end{exa}

\begin{rem}\label{rem2}
As suggested by a referee, to tackle with functional data with phase variation, one can first perform a curve registration \citep[e.g., the Fisher-Rao registration,][]{Srivastava2011} and then classify the resulting warping functions using standard functional classifiers. In our numerical examples, we show that this strategy does not evidently improve classification accuracy for functional data with phase variation and often worsens classification accuracy for those without phase variation; see Section E of the Appendix for details.
\end{rem}

\subsection{Main Method}\label{sc_method}
We see from the examples in Section~\ref{sc_model} that classification on manifold data can be sufficiently complicated especially under the functional data setting, even if the intrinsic dimension is low and the intrinsic class boundary is simple. To tackle the challenge, our aim is to learn low-dimensional representations of $\{X_i\}_{i=1}^n$, also termed manifold learning outcomes, that 
\begin{itemize}
    \item[(1)] maintain the unknown manifold structure from a certain aspect, and
    \item[(2)] separate the individuals according to their classes.
\end{itemize}
It is expected that a better classifier can be constructed based on these low-dimensional representations. The procedure with target (1) alone is termed unsupervised manifold learning, while that with both (1) and (2) is known as supervised manifold learning. It is important to note that target (2) alone is insufficient for classification purposes. That is, the classification model trained on manifold learning outcomes would not work well, if the individuals from different classes are separated as much as possible but the geometric structure is significantly compromised. This is because, to predict the label for a future observation, we need to construct a smooth interpolator from the functional space $L^2(\mathcal{T})$ to the manifold learning outcome space $\mathbb{R}^d$, which performs well only if the target (1) is properly achieved. To illustrate this argument, consider two adjacent points $X_1$ and $X_2$ with $Y_1\neq Y_2$. If the corresponding manifold learning outcomes $\tilde{Z}_1$ and $\tilde{Z}_2$ are far away, then it is impossible to accurately predict the manifold learning outcome of $X_0$ that is close to $X_1$ and $X_2$; see Section~\ref{sc_theo} for theoretical justification. Moreover, two- or three-dimensional manifold learning outcomes that ignore the target (1) render data visualization less useful, because they do not reveal the manifold structure of the covariate. Therefore, we intend to achieve a proper balance between these two targets: The desired manifold learning outcomes gather or disperse according to their classes; meanwhile, the particular manifold structure is maintained as well as possible. 

A basic idea in classification is that close or similar individuals should be assigned into the same class. A suitable metric to describe the proximity of two points in a manifold is the geodesic distance $d_\mathcal{M}$. Therefore, the aspect of the manifold structure that we intend to preserve is the geodesic distance, which is the same as the idea of the Isomap \citep{Tenenbaum2000}. Given pairwise geodesic distances of $\{X_i\}_{i=1}^n$, low-dimensional representations $\{\tilde{Z}_i\}_{i=1}^n\subset \mathbb{R}^d$ are obtained by applying the multidimensional scaling (MDS), where the Euclidean distances among the $\tilde{Z}_i$'s approximate the corresponding geodesic distances among the $X_i$'s. Roughly speaking, $\{\tilde{Z}_i\}_{i=1}^n$ represents a discrete representation of the unwrapped $\mathcal{M}$ embedded into $\mathbb{R}^d$.  

However, due to the dimension reduction from $\{X_i\}_{i=1}^n$ to $\{\tilde{Z}_i\}_{i=1}^n$ and estimation errors from a finite sample, the (estimated) distribution supports of different classes of $\{\tilde{Z}_i\}_{i=1}^n$ in $\mathbb{R}^d$ may be nonlinearly entangled, even if the underlying true distributions of different classes of $X$ in $\mathcal{M}$ are easily distinguishable; see Example~\ref{exa2}. In such scenarios, a classifier trained on $\{\tilde{Z}_i\}_{i=1}^n$ fails to accurately approximate the optimal classifier on $\mathcal{M}$. Therefore, we propose penalizing the geodesic distances of individuals of different classes. Specifically, letting  $\mathbb{D}_\xi$ be the $n\times n$ proximity matrix, we define its $(i,j)$th-entry as 
\begin{align}\label{eq_def_D}
\mathbb{D}_{\xi}(i,j) = d_\mathcal{M}(X_i,X_j)+\dfrac{\xi\cdot\mathds{1}\{Y_i\neq Y_j\} }{d_\mathcal{M}(X_i,X_j)+\sqrt{\xi}}\,,
\end{align}
where $\mathds{1}\{\cdot \}$ is the indicator function, and $\xi>0$ is a tuning parameter. That is, for two individuals from different classes, their geodesic distance is penalized by adding $\xi/\{d_\mathcal{M}(X_i,X_j)+\sqrt{\xi}\}$. The motivation for using $\xi/\{d_\mathcal{M}(X_i,X_j)+\sqrt{\xi}\}$ is twofold. First, the penalty is negatively correlated with $d_\mathcal{M}(X_i,X_j)$. This is desirable, because for two points with large $d_\mathcal{M}(X_i,X_j)$, their classes being different can be largely quantified by their geodesic distance, and there is no need to penalize further on their proximity. Second, the proximity rank for individuals from different classes remains unchanged, i.e., $\mathbb{D}_{\xi}(i,j)<\mathbb{D}_{\xi}(i,k)$ if $d_{\mathcal{M}}(X_i,X_j)< d_{\mathcal{M}}(X_i,X_k)$, for $Y_i\neq Y_j$ and $Y_i\neq Y_k$. These two desired features of $\xi/\{d_\mathcal{M}(X_i,X_j)+\sqrt{\xi}\}$ properly meet the aforementioned target (1), i.e., alleviating rupture of the original manifold structure (in particular, pairwise geodesic distances). To ensure the target (2), we may simply set a larger $\xi$. The penalized geodesic distance defined in~\eqref{eq_def_D} distinguishes our approach from the supervised Isomap methods in the literature \citep{Vlachos2002,Geng2005,Zhang2018} and may be of interest in its own right.  

By applying the MDS on $\mathbb{D}_\xi$, we obtain the supervised manifold learning outcomes $\{Z_i\}_{i=1}^n\subset \mathbb{R}^d$, which are regarded as low-dimensional representations of the functional data that include the label information. In addition, they can be used for visualization if we set $d=2$ or $3$. We assume that $\mathcal{M}$ is characterized by a single coordinate map $\mu:\mathcal{M}\to \mathcal{U}$, where $\mu$ is a diffeomorphism (i.e., a one-to-one differentiable map whose inverse is also differentiable) and $\mathcal{U}$ is an open set in $\mathbb{R}^d$. The assumption that $\mathcal{M}$ is covered by a single chart can be restrictive, but is routinely assumed in the literature of manifold learning for tractable theoretical investigation~\citep{Tenenbaum2000,Zhang2004,Chen2012}. The outcomes $\{Z_i\}_{i=1}^n$ provide an estimator of $\mu$ at $\{X_i\}_{i=1}^n$. To construct a functional classifier for future observations, we need to develop a global estimator $\tilde{\mu}$ of $\mu$, which can be achieved by regressing $\{Z_i\}_{i=1}^n$ on $\{X_i\}_{i=1}^n$ using a nonparametric smoothing technique. The procedure is also known as the out-of-sample interpolation in the literature on manifold learning. Finally, our proposed functional classifier is defined as $H(\cdot)=H_d\{\tilde{\mu}(\cdot)\}$, where $H_d:\mathbb{R}^d\to \mathcal{Y}$ is any suitable classifier trained on $\{Z_i,Y_i\}_{i=1}^n$. Specifically, we prefer simple geometry-driven classifiers for constructing $H_d$, such as the $k$-NN classifier and the SVM, because we expect that the optimal boundaries for separating classes in the manifold can be well unfolded onto $\mathbb{R}^d$.

\section{Estimation}\label{sc_est}
We introduce a detailed estimation procedure to implement the proposed methodology as in Section~\ref{sc_method}. We recover functional data from discrete observations in Section~\ref{sc_est_fd}, estimate the geodesic distance in Section~\ref{sc_est_gd}, introduce the interpolation and classification based on the manifold learning outcomes in Section~\ref{sc_est_cl}, and finally discuss the selection of the tuning parameters in Section~\ref{sc_est_tp}.

\subsection{Functional Data Estimation}\label{sc_est_fd}

In practice, we often do not directly observe $X_i$ but its discrete and noisy version $\{(T_{ij},\tilde{X}_{ij})\}_{j=1}^{J_i}$, where 
\begin{align}\label{eq_X_obs}
\tilde{X}_{ij} = X_{i}(T_{ij}) + \epsilon_{ij}\,.
\end{align}
Here, $T_{ij} \in \mathcal{T}$ are the time points of observations and $\epsilon_{ij}$ are the random errors with $\bE(\epsilon_{ij})=0$ and $\mathrm{Var} (\epsilon_{ij})< \infty$. To recover smooth trajectories from discrete and noisy observations, when the sampling is not too sparse, i.e., $\min_{i} J_{i}$ is not too small and the $T_{ij}$'s roughly ranges across the whole interval $\mathcal{T}$, we apply the ridged local linear estimator \citep{Fan1996,Lin2021} on individual observations $\{(T_{ij},\tilde{X}_{ij})\}_{j=1}^{J_i}$ to obtain the estimator $\hat{X}_i$ of $X_i$. Specifically, the standard local linear estimator of $X_i(t)$ is given by $(T_0S_2-T_1S_1)/(S_0S_2-S_1^2)$, where
\begin{align*}
S_k = \dfrac{1}{J_i} \sum_{j=1}^{J_i} K\Big(\dfrac{T_{ij}-t}{h_i}\Big) \Big( \dfrac{T_{ij}-t}{h_i}\Big)^k\,,~T_k=\dfrac{1}{J_i} \sum_{j=1}^{J_i} K\Big(\dfrac{T_{ij}-t}{h_i}\Big)\Big( \dfrac{T_{ij}-t}{h_i}\Big)^k \tilde{X}_{ij}\,,
\end{align*}
for $k=0,1$ and $2$. Here, $K$ is the kernel and $h_i>0$ the bandwidth. The kernel $K$ is often a symmetric density function, e.g., the standard Gaussian density, and the bandwidth $h_i$ can be chosen by the cross-validation or plug-in criteria. To prevent pathological cases where the denominator $(S_0S_2-S_1^2)$ is near zero, we define the ridged local linear estimator of $X_i(t)$ as 
\begin{align}\label{eq_X_est}
\hat{X}_i(t) = \dfrac{T_0S_2-T_1S_1}{S_0S_2-S_1^2+\lambda \sign(S_0S_2-S_1^2)\mathds{1}\{|S_0S_2-S_1^2|<\lambda\}}\,, ~~\textrm{for } t \in \mathcal{T}\,,
\end{align}
where the ridge parameter $\lambda$ can be set to $J_i^{-2}$ following \citet{Lin2021}. On the other hand, when the sampling is sparse or the data are partially observed, other techniques \citep[e.g.,][]{Yao2005,Kneip2020} can be applied to obtain the smooth trajectories $\hat{X}_i$. For the sake of brevity, our theoretical analysis only concerns the estimator defined in~\eqref{eq_X_est}.

\subsection{Geodesic Distance Estimation}\label{sc_est_gd}
To estimate the geodesic distance, we employ the functional parallel transport unfolding (FPTU) developed in \citet{Tan2024}, which is shown to be more accurate and robust than the traditional FIsomap. The first step is to construct an $n\times n $ weighting matrix $G$, where $G(i,j)=\|\hat{X}_i-\hat{X}_j\|_{L^2}$ if $\hat{X}_i$ and $\hat{X}_j$ are adjacent, and $G(i,j)=\infty$ otherwise. To ensure that the graph induced by $G$ is connected so that the final manifold learning outcomes can be embedded into the same coordinate system, we utilize the minimum spanning tree idea in \citet{Dimeglio2014} to define the adjacency of $\{\hat{X}_i\}_{i=1}^n$; see Section B of the Appendix for details. For individuals $i$ and $j$, we define the geodesic path as the shortest path on $G$ between them, obtained by applying Dijkstra’s algorithm on $G$. The FIsomap uses the length of the geodesic path to estimate the geodesic distance, while the FPTU estimates the geodesic distance by the length of the unfolded geodesic path, which is obtained through the discrete parallel transport as follows.

We need to estimate the tangent space at each data point to perform the discrete parallel transport. The tangent space at a given point on the manifold serves as the best linear approximation of the manifold locally around that point. As PCA yields a linear projection of the data that is optimal in the sense of mean squared error, a natural idea of estimating tangent spaces is to utilize the \emph{local} PCA \citep{Singer2012}. This reasoning also applies in the context of functional data. Therefore, to estimate $T_i\mathcal{M}\equiv T_{X_i}\mathcal{M}$, the tangent space at $X_i$, we consider the neighborhood $\mathcal{N}_{i,\rm{PCA}}$ that contains $k_{\mathrm{PCA}}$-nearest neighbors ($k_{\mathrm{PCA}}$-NN) of $\hat{X}_i$ (note that $X_i$ is not available). We define the local empirical covariance function around $\hat{X}_i$ as
\begin{align*}
\hat{\Gamma}_i(s,t) = \dfrac{1}{k_{\rm{PCA}}} \sum_{j\in \mathcal{N}_{i,\rm{PCA}}}\{\hat{X}_j(s)-\hat{\mu}_i(s)\} \{\hat{X}_j(t)-\hat{\mu}_i(t)\}\,,~~\textrm{for } s,t\in \mathcal{T}\,,
\end{align*}
where $\hat{\mu}_{i}(t) = \sum_{j\in \mathcal{N}_{i,\rm{PCA}}} \hat{X}_j/k_{\rm{PCA}}$ is the empirical local mean around $\hat{X}_i$. The estimator of $T_i \mathcal{M}$ is then defined as $\hat{T}_i \mathcal{M} =\mathrm{span}\{\hat{\phi}_{i1},\ldots,\hat{\phi}_{id}\}$, where $\hat{\phi}_{i1},\ldots,\hat{\phi}_{id}$ are the first $d$ eigenfunctions of $\hat{\Gamma}_i$.  If the intrinsic dimension $d$ is unknown, we adapt the minimal neighborhood method \citep{Facco2017} to estimate it; see Section B of the Appendix for details.

Let $\hat{\Phi}_{ij}$ denote a $d\times d$ matrix with the $(k,s)$th entry $\hat{\Phi}_{ij}(k,s) = \langle\hat{\phi}_{ik},\hat{\phi}_{js} \rangle$, where $\langle\cdot,\cdot \rangle$ is the inner product in $L^2(\mathcal{T})$. It can be shown that $R_{j,i}=VU^\top$ provides an approximate parallel transport from $\hat{T}_i \mathcal{M}$ to $\hat{T}_j \mathcal{M}$, where $\hat{\Phi}_{ij}=U\Sigma V^\top$ is the singular value decomposition of $\hat{\Phi}_{ij}$ \citep{Tan2024}. This means that, for a vector $u=(u_1,\ldots,u_d)^\top \in \hat{T}_i \mathcal{M}$ in terms of the basis $\{\hat{\phi}_{i1},\ldots,\hat{\phi}_{id}\}$, the vector $\hat{R}_{j,i}(u_1,\ldots,u_d)^\top$ is the parallel transported $u$ in $\hat{T}_j \mathcal{M}$ under the basis $\hat{\Phi}_j$. For a geodesic path $(i=i_0,i_1,\ldots,i_{m-1},i_m=j)$ in $G$, we parallel transport all the edges on the path to the last tangent space $\hat{T}_j \mathcal{M}$ and then aggregate them, which yields the unfolded geodesic. The Euclidean norm of the unfolded geodesic provides a robust approximation of the geodesic distance between $i$ and $j$. Specifically, we first project $\hat{V}_{i_k} = \hat{X}_{i_{k-1}}-\hat{X}_{i_k}$ to the tangent space $\hat{T}_{i_k} \mathcal{M}$, for $k=1,\ldots,m$. Let $v_{i_k} = (\langle \hat{V}_{i_k},\hat{\phi}_{i_{k}1}\rangle,\ldots,\langle \hat{V}_{i_k},\hat{\phi}_{i_{k}d}\rangle)^\top$ be the coordinates of the projected $\hat{V}_{i_k}$. It follows that $v_{i_k,m} =  \hat{R}_{i_m,i_{m-1}}\cdots \hat{R}_{i_{k+1},i_k}  v_{i_k}$ is the parallel transported $v_{i_k}$ in $\hat{T}_{j}\mathcal{M}$, for $k=1,\ldots,m-1$. For $k=m$, we define $v_{i_k,m} = v_{j,m}  = v_j$, which is already in $\hat{T}_j \mathcal{M}$ and no parallel transport is needed. Finally, the aggregated vector $v_{i,m} = \sum_{k=1}^m v_{i_k,m}$ in $\hat{T}_j \mathcal{M}$ can be seen as the unfolded geodesic from $\hat{X}_i$ to $\hat{X}_j$. As the Euclidean norm of the unfolded geodesic from $\hat{X}_j$ to $\hat{X}_i$ is generally different from that of $v_{i,m}$, we compute the average of these two as $\hat{d}_\mathcal{M}(X_i,X_j)$, our estimator of $d_\mathcal{M}(X_i,X_j)$.

\subsection{Interpolation and Classification}\label{sc_est_cl}
The estimated proximity matrix $\hat{\mathbb{D}}_\xi$ is obtained by plugging $\hat{d}_\mathcal{M}(X_i,X_j)$ into~\eqref{eq_def_D}, and the corresponding manifold learning outcomes $\{\hat{Z}_i\}_{i=1}^n \subset \mathbb{R}^d$ are obtained by applying the MDS to $\hat{\mathbb{D}}_\xi$. Recalling that $\mathcal{M}$ is assumed to be characterized by a coordinate map $\mu:\mathcal{M}\to \mathcal{U}$, we need to estimate $\mu$ globally to construct the functional classifier. The low-dimensional representations $\hat{Z}_i$ provide estimates of $\mu$ at the sample points $X_i$.
To construct an estimator of $\mu$ at any given point $x\in L^2(\mathcal{T})$,
we utilize the local linear regression on the tangent space \citep{Lin2021}. 

Specifically, the tangent space $T_x \mathcal{M}$ can be estimated following the method in~Section~\ref{sc_est_gd}. That is, we define $\hat{T}_x \mathcal{M}=\mathrm{span}\{\hat{\phi}_{x1},\ldots,\hat{\phi}_{xd}\}$, where $\{\hat{\phi}_{x1},\ldots,\hat{\phi}_{xd}\}$ are the first $d$ eigenfunctions of 
\begin{align*}
\hat{\Gamma}_x(s,t) = \dfrac{1}{k_{\rm{PCA}}} \sum_{j\in \mathcal{N}_{x,\rm{PCA}}}\{\hat{X}_j(s)-\hat{m}_x(s)\} \{\hat{X}_j(t)-\hat{m}_x(t)\}\,,~~\textrm{for } s,t\in \mathcal{T}.
\end{align*}
Here, $\hat{m}_x(t) = \sum_{j\in \mathcal{N}_{x,\rm{PCA}}} \hat{X}_j/k_{\rm{PCA}}$ and $\mathcal{N}_{x,\rm{PCA}}$ contains $k_{\mathrm{PCA}}$-NN of $x$. We project $\hat{X}_i$ onto $\hat{T}_x \mathcal{M}$, whose coordinates in terms of $\{\hat{\phi}_{x1},\ldots,\hat{\phi}_{xd}\}$ are $\hat{\chi}_i = (\langle\hat{X}_i-x,\hat{\phi}_{x1}\rangle,\ldots, \langle\hat{X}_i-x,\hat{\phi}_{xd}\rangle)^\top$, for $i=1,\ldots,n$. The estimator of $\mu(x)$ is defined as 
\begin{align}\label{eq_mu_est}
\hat{\mu}(x)^\top = e_1^\top (\hat{\mathbb{X}}^\top \mathbb{W} \hat{\mathbb{X}})^{-1}\hat{\mathbb{X}}^\top \mathbb{W} \hat{\mathbb{Z}}\,,
\end{align}
where $\mathbb{W} = \mathrm{diag}\big(K_{h_{\mathrm{reg}}}(\|\hat{X}_1-x \|_{L^2}),\ldots,K_{h_{\mathrm{reg}}}(\|\hat{X}_n-x \|_{L^2})\big)$ with $K_h = K(\cdot/h)/h^d$, $\hat{\mathbb{Z}}=(\hat{Z}_1,\ldots,\hat{Z}_n)^\top$, $e_1=(1,0,\ldots,0)^\top$ is a $d+1$-dimensional vector, and
\begin{align*}
\hat{\mathbb{X}}= 
\begin{pmatrix}
1 &\ldots & 1 \\
\hat{\chi}_1&\ldots & \hat{\chi}_n
\end{pmatrix}^\top\,.
\end{align*}
In the case that $\hat{\mathbb{X}}^\top \mathbb{W} \hat{\mathbb{X}}$ in~\eqref{eq_mu_est} is nearly singular, which may happen when a too small bandwidth is used, a ridge term similar to~\eqref{eq_X_est} can be used to stabilize computation: We replace $\hat{\mathbb{X}}^\top \mathbb{W} \hat{\mathbb{X}}$ by $\hat{\mathbb{X}}^\top \mathbb{W} \hat{\mathbb{X}}+\lambda_n I_{d+1}$, where $I_{d+1}$ is an identity matrix of $d+1$ dimensions and $\lambda_n$ is a small positive number, say $n^{-3}$.
 
Finally, a classical multivariate classifier $\hat{H}_d:\mathbb{R}^d\to \mathcal{Y}$, say the $k$-NN classifier, is trained on $\{\hat{Z}_i,Y_i\}_{i=1}^n$, and our functional classifier is defined as $\hat{H}(\cdot)=\hat{H}_d\{\hat{\mu}(\cdot)\}$.

\subsection{Tuning Parameters Selection}\label{sc_est_tp}
Our procedure involves selecting several tuning parameters. In the first step of recovering functional data, we suggest using the plug-in bandwidth~\citep{Ruppert1995} for the ridged local linear estimator. Regarding geodesic distance estimation, following the suggestion of \citet{Tan2024}, one may experiment several values around $n^{2/(d+2)}$ for $k_{\rm{PCA}}$. If the purpose is dimension reduction solely, one may test a few values for $\xi$ and apply the MDS on $\mathbb{D}_\xi$ to obtain the low-dimensional representations.

If the purpose is classification, it remains to choose $\xi$ in computing $\mathbb{D}_\xi$, which significantly affects the manifold learning outcomes, and the functional regression bandwidth $h_{\rm{reg}}$, which determines the ability of the out-of-sample interpolation. These two parameters affect each other and are both crucial to classification performance. Therefore, a careful selection for these two parameters is needed. We propose a nested $L$-fold cross-validation (CV) procedure to select them. Specifically, we first estimate pairwise geodesic distances using the whole dataset. We then randomly split the dataset into $L$ parts denoted by $\mathcal{F}_{1},\ldots,\mathcal{F}_{L}$. Let $\mathcal{F}_{-\ell}$ denote the remaining dataset with $\mathcal{F}_{\ell}$ excluded. For a given $\mathcal{F}_{-\ell}$, we further randomly split it into $L$ parts, $\mathcal{S}_{1}^\ell,\ldots,\mathcal{S}_{L}^\ell$, and denote by $\mathcal{S}_{-m}^\ell$ the remaining dataset with $\mathcal{S}_{m}^\ell$ excluded. The outer split samples $\mathcal{F}_\ell$ are used to select $\xi$, while the inner split samples $\mathcal{S}^\ell_m$ are used to select $h_{\rm{reg}}$. Given a classifier $\hat{H}_d$ on $\mathbb{R}^d$, we define the CV loss for $\xi$ as
\begin{align*}
    \mathrm{CV}_{\mathcal{F}} (\xi) = \sum_{\ell=1}^L\sum_{i \in \mathcal{F}_\ell} \mathds{1}\big(Y_i\neq \hat{H}_{d,-\ell}\{\hat{\mu}_{-\ell,h_{\rm{reg},\ell}^{\rm{CV}}}(\hat{X}_i) \}\big)\,,
\end{align*}
where $\mathds{1}(\cdot)$ denotes the indicator function, $\hat{H}_{d,-\ell}$ is obtained by training the classifier on $\{\hat{Z}_i,Y_i\}_{i\in \mathcal{F}_{-\ell}}$, and $\hat{\mu}_{-\ell,h_{\rm{reg},\ell}^{\rm{CV}}}$ is the estimated coordinate map using $\{\hat{X}_i,\hat{Z}_i\}_{i\in \mathcal{F}_{-\ell}}$ and $h_{\rm{reg},\ell}^{\rm{CV}}$. The CV bandwidth $h_{\rm{reg},\ell}^{\rm{CV}}$ is defined as the minimum of the CV loss for $h_{\rm{reg}}$,
\begin{align*}
    \mathrm{CV}_{\mathcal{S}^\ell}(h) = \sum_{m=1}^L \sum_{i\in \mathcal{S}_m^\ell} \|\hat{Z}_i-\hat{\mu}_{-m,h}^\ell(\hat{X}_i)\|_{\mathbb{R}^d}^2\,,
\end{align*}
where $\|\cdot\|_{\mathbb{R}^d}$ denotes the Euclidean distance in $\mathbb{R}^d$, and $\hat{\mu}_{-m,h}^\ell$ is the coordinate map trained on $\{\hat{X}_i,\hat{Z}_i\}_{i\in \mathcal{S}_{-m}^\ell}$ with bandwidth $h$. The pair $(\xi^{\rm{CV}},\bar{h}_{\rm{reg}}^{\rm{CV}})$ with $\bar{h}_{\rm{reg}}^{\rm{CV}} = \sum_{\ell=1}^L h_{\rm{reg},\ell}^{\rm{CV}}/L $ such that $\xi^{\rm{CV}}=\argmin_\xi  \mathrm{CV}_{\mathcal{F}} (\xi)$ is our final choice for $\xi$ and $h_{\rm{reg}}$.

\section{Theoretical Properties}\label{sc_theo}
Recall that we assume that the manifold $\mathcal{M}$ is characterized by a single coordinate map $\mu:\mathcal{M} \to \mathcal{U}$ with an open $\mathcal{U} \in \mathbb{R}^d$. The coordinate map is not unique: For any diffeomorphism $\psi$, $\psi\{\mu(\cdot)\}$ is also a coordinate map. Since our manifold learning procedure aims to preserve the geodesic distance (up to a penalty for different labels), we are interested in coordinate maps that preserve the distance. Therefore, we assume that there exists a coordinate map $\mu$ that is isometric, i.e., $\|\mu(x_1)-\mu(x_2)\|_{\mathbb{R}^d}=d_g(x_1,x_2)$ for any $x_1,x_2\in\mathcal{M}$. If such a $\mu$ exists, it is still not unique, because a rigid transformation applied to $\mu$ does not change the Euclidean distance. In addition, identifying $\mu$ globally from the observed data alone without any assumptions is not possible, because the low-dimensional representations $Z_i=\mu(X_i)$ are not observed. For $i=1,\ldots,n$, let $\hat{\mu}(X_i)=\hat{Z}_i$ denote the manifold learning outcome. To avoid this dilemma, \citet{Chen2012} assumed beforehand that $E\{\sup_{i=1,\ldots,n}\|\hat{\mu}(X_i)-\mu(X_i)\|_{\mathbb{R}^d}\}\to 0$ at a given rate, and investigated the convergence rate of their $\hat{\mu}^{-1}$ at any given point. In contrast, we show that a more concrete convergence rate of $\sup_{i=1,\ldots,n}\|\hat{\mu}(X_i)-\mu(X_i)\|_{\mathbb{R}^d}$ can be obtained under suitable conditions. All the proofs of the theoretical results can be found in Sections~C and D of the Appendix.

Our key assumption for identifying $\mu$ from the observed data is as follows.
\begin{itemize}[wide] 
    \item[\emph{Assumption} 1.] There exists a $c>0$ such that
    \begin{align}\label{eq_assum}
    \sup_{i=1,\ldots,n} \|\hat{\mu} (X_i)-\mu(X_i) \|_{\mathbb{R}^d} \leq c\cdot\sup_{i,j=1,\ldots,n} 
 |\|\hat{\mu} (X_i)-\hat{\mu}(X_j) \|_{\mathbb{R}^d}-\|\mu (X_i)-\mu(X_j) \|_{\mathbb{R}^d}|\,.
    \end{align} 
\end{itemize}
The assumption above requires that the distance between $\hat{\mu}$ and $\mu$ at the sample points can be controlled by the difference of with-in-sample distances $\|\hat{\mu} (X_i)-\hat{\mu}(X_j) \|_{\mathbb{R}^d}$ and $\|\mu (X_i)-\mu(X_j) \|_{\mathbb{R}^d}$. If $\mu$ satisfies Assumption 1, $\psi\{\mu(\cdot)\}$ no longer satisfies this assumption with a rigid transformation $\psi$ on $\mathbb{R}^d$. Therefore, Assumption 1 can be seen as an identifiability condition for $\mu$. The idea of imposing such an assumption is that the convergence rate of the right hand side of~\eqref{eq_assum} is tractable. It is not necessary if the convergence rate of the left hand side of~\eqref{eq_assum} is given.

Recall that the manifold learning outcomes $\hat{Z}_i=\hat{\mu}(X_i)$ are obtained by applying the MDS on $\hat{\mathbb{D}}_\xi$. The MDS may not exactly preserve each pair of distances. Let $\epsilon_{\rm{MDS}}=\sup_{i=1,\ldots,n}|\hat{\mathbb{D}}_\xi(i,j)-\|\hat{Z}_i-\hat{Z}_j\|_{\mathbb{R}^d}|$ to quantify the discrepancy. The following result provides the with-in-sample convergence rate of $\hat{\mu}$.

\begin{prop}\label{prop1}
Under Assumption 1 and Conditions (A1) to (A4), (B1) to (B5) in Section~C.1 of the Appendix, assuming that $\xi,\epsilon_{\rm{MDS}}\to 0$, and the number of vertices of any geodesic path on $G$ is uniformly bounded by $m$, we have
\begin{align*}
\sup_{i=1,\ldots,n} \|\hat{\mu} (X_i)-\mu(X_i) \|_{\mathbb{R}^d}=O_p\{\xi^{1/2} +\epsilon_{\rm{MDS}}+m^2(h_{\rm{PCA}} + h_g^3 + h_g h_d^2\kappa_s)\}\,,
\end{align*}
where $h_{\rm{PCA}},h_g,h_d$ and $\kappa_s$ are given in Section C.1 of the Appendix.
\end{prop}

We see from Proposition~\ref{prop1} that the convergence rate of $\hat{\mu}$ on the sample points consists of three parts: The convergence rate of the geodesic distance estimation $m^2(h_{\rm{PCA}} + h_g^3 + h_g h_d^2\kappa_s)$ that is derived in \citet{Tan2024}, the loss of MDS $\epsilon_{\rm{MDS}}$, and the tuning parameter $\xi$. It is necessary to require $\xi\to 0$, because the MDS mapping is otherwise not continuous. This agrees with our discussion that the manifold structure should be maintained as well as possible in Section~\ref{sc_method}. The assumption that $\epsilon_{\rm{MDS}}\to 0$ is also mild, because $\epsilon_{\rm{MDS}}$ can be made arbitrarily small in practice by selecting a sufficiently large embedded dimension $d$. 

Next, we provide the out-of-sample convergence rate of $\hat{\mu}$.
\begin{prop}\label{prop2}
Under the same conditions as in Proposition~\ref{prop1} and Conditions (C1) and (C2) in Section~C.2 of the Appendix, suppose  $h_{\rm{reg}}>h_{\rm{PCA}}$, for any $\epsilon>0$ and any given point $x\in\mathcal{M}$, we have
\begin{align*}
\|\hat{\mu}(x)-\mu(x)\|_{\mathbb{R}^d} = O_p\{h_{\rm{reg}}^2+h_{\rm{reg}}^{-\alpha-\epsilon d}J^{-\beta}+\xi^{1/2} +\epsilon_{\rm{MDS}}+m^2(h_{\rm{PCA}} + h_g^3 + h_g h_d^2\kappa_s)\}\,,
\end{align*}
where $\alpha=1$ if $x$ is at least $h_{\rm{reg}}$ away from the boundary of $\mathcal{M}$, and $\alpha=2$ otherwise. 
\end{prop}
Compared to Proposition~\ref{prop1}, the convergence rate of $\hat{\mu}$ at a given point includes the additional terms $h_{\rm{reg}}^2+h_{\rm{reg}}^{-\alpha-\epsilon d}J^{-\beta} $, which are the convergence rate of the bias of the local linear estimator in the tangent space~\citep{Lin2021}. Unlike the regression setting where the outcome variable is observed with error, the outcomes here are obtained by the supervised manifold learning procedure, and thus there is no variance component. 

Propositions~\ref{prop1} and \ref{prop2} provide convergence rates of the estimated coordinate map, but, typically, one is more interested in the classification properties induced by the coordinate map. However, such properties depend on the specific multivariate classifier applied on the manifold learning outcomes. To investigate the theoretical classification properties of our method, we focus on the vanilla $k$-NN classifier with binary outcomes $Y\in\{0,1\}$ as an illustration. Specifically, for any new observation $X_0\in \mathcal{M}$, we first obtain the manifold representation $\hat{Z}_0=\hat{\mu}(X_0)$ and reorder the data $\{X_{(i)},Y_{(i)}\}_{i=1}^n$ according to $\|\hat{Z}_{(1)}-\hat{Z}_0\|_{\mathbb{R}^d}\leq ...\leq \|\hat{Z}_{(n)}-\hat{Z}_0\|_{\mathbb{R}^d}$. The $k$-NN classifier is then defined as 
\begin{align}\label{eq_knn}
\hat{H}_{k-\rm{NN}}(X_0)=\left\{
\begin{aligned}
&0, ~\textrm{if }~ \frac{1}{k}\sum_{i=1}^k Y_{(i)} \leq \frac{1}{2}\,,\\
&1,~\textrm{otherwise\,.}
\end{aligned}
\right.
\end{align}

Let $H^*(X)=\mathds{1}\{\eta(X)>1/2\}$ denote the Bayes classifier, where $\mathds{1}(\cdot)$ is the indicator function, and $\eta(x)=E(Y|X=x)$. It is well known that $H^*$ is optimal in the sense that it achieves the smallest misclassification error. That is, for any classifier $H:L^2(\mathcal{T})\to \mathcal{Y} $, $\mathcal{R}(H^*)\leq \mathcal{R}(H)$, where $\mathcal{R}(H)=P(H(X)\neq Y)$, also referred to as the misclassification risk. The following theorem shows that our functional classifier coupled with the $k$-NN classifier is asymptotically optimal.

\begin{theo}\label{theo1}
Under the same conditions as Proposition~\ref{prop2} and Conditions~(D1) to (D3) in Section D of the Appendix, for all $\delta>0$, there exists $N_\delta\in \mathbb{N}$, such that if $n\gtrsim \max\{N_\delta,\delta^{-(d+2)/d}\}$, $k=\lfloor \delta^{-2/d} \rfloor$, then
\begin{align}\label{eq_R_kNN}
\mathcal{R}(\hat{H}_{k-\rm{NN}})-\mathcal{R}(H^*) \leq C \delta^{(1+\gamma)/d}\,,
\end{align}
where $C>0$ is a finite constant and $\gamma$ is specified in Conditions (D2).
\end{theo}
Theorem~\ref{theo1} shows that the risk between $\hat{H}_{k-\rm{NN}}$
and $H^*$ can be arbitrarily small, i.e., $\hat{H}_{k-\rm{NN}}$ is asymptotically optimal. The value $N_\delta$ is related to the convergence rate of $\hat{\mu}$ as in Proposition~\ref{prop2}. However, a more specific form for $N_\delta$ is not available, because Proposition~\ref{prop2} is a statement of the convergence in probability rather than the almost sure convergence; see the proof for more details. If $N_\delta>\delta^{-(d+2)/d}$, then $\delta^{-(d+2)/d}=o(n)$, which implies that $n^{-(1+\gamma)/(d+2)}/\delta^{(1+\gamma)/d}=o(1)$. That is, the convergence rate of the bound in~\eqref{eq_R_kNN} is slower than $n^{-(1+\gamma)/(d+2)}$ provided that $N_\delta>\delta^{-(d+2)/d}$. If $N_\delta<\delta^{-(d+2)/d}$ otherwise, we can in fact choose $n\asymp \delta^{-(d+2)/d}$, which implies that $\mathcal{R}(\hat{H}_{k-\rm{NN}})-\mathcal{R}(H^*) \leq C n^{-(1+\gamma)/(d+2)}$, the minimax optimal rate of the $k$-NN classifier as shown in \citet{Gadat2016}.  

\section{Simulation Studies}\label{sc_sim}
We conduct simulation experiments to evaluate performance of the functional supervised manifold learning (FSML) procedure on finite samples. To construct a functional classifier, we combine the FSML with three classical classifiers: the $k$-NN classifier (FSML$_{k-\rm{NN}}$), the SVM (FSML$_{\rm{SVM}}$), and the linear discriminant analysis (FSML$_{\rm{LDA}}$). For comparison, we consider several existing functional classifiers: the centroid classifier~\citep[CC,][]{Delaigle2012}, the functional quadratic discriminant analysis~\citep[FQDA,][]{Delaigle2013,Galeano2015}, the nonparametric Bayesian classifier~\citep[NB,][]{Dai2017} based on the kernel density estimate, and the functional deep neural network classifier~\citep[FDNN,][]{Wang2023}. The code to implement the FDNN is available on the authors' personal website, while the other methods are reproduced using our code. 

We generate data from a few functional manifold models as well as two Gaussian process models:
\begin{enumerate}
    \item[(i)]  Time warping: $X_i(t) = Z_{i1}\mu\{\gamma_i(t)\}$, where $\mu(t) = \phi_{0.2,0.08}(t)+\phi_{0.5,0.1}(t)+\phi_{0.8,0.13}(t)$, $\gamma_i(t) = \{\exp(Z_{i2}t)-1\}/\{\exp(t)-1\}$ if $Z_{i2}\neq 0$ and $\gamma_i(t) =t  $ if $Z_{i2}=0$. Here $Z_{i1}\overset{iid}{\sim}{\textrm{Gamma}}(a=4,b=0.5)$, $Z_{i2}|Y_i=0\overset{iid}{\sim} U(-1,0.2), Z_{i2}|Y_i=1\overset{iid}{\sim} U(-0.2,1)$, and $\phi_{\mu,\sigma}$ denotes the probability density function of $N(\mu,\sigma^2)$. 

    \item[(ii)] Two Swiss rolls: $(X_i(t)|Y_i) = Z_{i1}\cos(Z_{i1}+\pi\mathds{1}\{Y_i=1\})\sin(2\pi t)+Z_{i1}\sin(Z_{i1}+\pi\mathds{1}\{Y_i=1\})\cos(2\pi t) + Z_{i2}\sin(4\pi t)$, where $Z_{i1}\overset{iid}{\sim} U(0,2\pi)$, $Z_{i2}\overset{iid}{\sim} U(0,8)$.

    \item[(iii)] Torus: $X_i(t)=\{2+\cos(\theta_i)\}\cos(\phi_i)\sin(2\pi t)+\{2+\cos(\theta_i)\}\sin (\phi_i) \cos(2\pi t)+\sin(\theta_i)\sin(4\pi t)$, where $(\theta_i,\phi_i|Y_i=0)$ follows a uniform distribution on $0<\theta<\phi<2\pi$, and $(\theta_i,\phi_i|Y_i=1)$ follows a uniform distribution on $0<\phi\leq \theta<2\pi$.

    \item[(iv)] Gaussian I: $X_i(t) = \xi_{i1}\log(t+2)+\xi_{i2}t+\xi_{i3}t^3$, where $(\xi_{i1},\xi_{i2},\xi_{i3})^\top|Y_i=k\overset{iid}{\sim}N(\mu_k,\Sigma_k)$ with $\mu_0=(-1,2,-3)^\top,\mu_1=(-1/2,5/2,-5/2)^\top,\Sigma_0^{1/2} = \diag(3/5,2/5,1/5)$ and $\Sigma_1^{1/2} = \diag(9/10,1/2,3/10)$.

    \item[(v)] Gaussian II: $X_i(t) = \mu_i(t) + \sum_{j=1}^{50} Z_{ij} \phi_j(t)$, where $\mu_i(t) = 0$ if $Y_i=0$ and $\mu_i(t)=1$ otherwise, $Z_{ij}|Y_i=k\sim N(0,\sigma_{jk}^2)$ with $\sigma_{j0} = \exp(-j/6)$ and $\sigma_{j1} = \exp(-j/4)$, $\phi_1(t)=1,\phi_{2\ell+1}(t)=\sqrt{2}\sin(2\ell\pi t)$, and $\phi_{2\ell}(t) = \sqrt{2}\cos(2\ell\pi t) $.
\end{enumerate}
Under all models above, the label variable $Y_i\overset{iid}{\sim}b (1,0.5)$ corresponding to a binary classification setting. The observed data are $(\tilde{X}_{ij},Y_i)$, where $\tilde{X}_{ij}=X_i(t_{j})+\epsilon_{ij}$, for $i=1,\ldots,n$ and $j=1,\ldots,J$. Here, $0=t_1< \ldots< t_J=1$ are equidistant points on $[0,1]$, and $\epsilon_{ij}\overset{iid}{\sim} N(0,\hat{V}_X/20)$, where $\hat{V}_X$ is the sample variance of $X$ integrated on $[0,1]$.

Model (i), modified from model (iii) in \citet{Tan2024}, is a time-warping model exhibiting significant phase variation, which is a classical example of a functional manifold~\citep{Srivastava2011,Chen2012}. Models (ii) and (iii) are two connected Swiss rolls (see the left panel of Figure~\ref{fg_exm1} for an illustration) and a torus embedded into a functional space, respectively. These geometric objects are two-dimensional manifold examples commonly used in the literature on manifold learning~\citep{Tenenbaum2000,Meila2024}. Model (iv) is a low-dimensional Gaussian model adopted from \citet{Wang2024}, while model (v) is a high-dimensional one adopted from \citet{Dai2017}.

\begin{table}[t]
\centering
\caption{Mean (standard deviation) misclassification errors in percentage under all simulation models with the best result highlighted in boldface.}\label{tb1}
\resizebox{\columnwidth}{!}{
%\begin{threeparttable}
\begin{tabular}{ccccccccc}
\hline
Model & $n$ & FSML$_{20-\rm{NN}}$ & FSML$_{\rm{SVM}}$ & FSML$_{\rm{LDA}}$ & CC & FQDA & NB & FDNN\\
\hline
\multirow{2}{*}{(i)} & 200 & 19.2 (1.9) & 19.1 (1.9) & 19.2 (2.0) & {\bf 17.6} (1.7) & 28.7 (3.7) & 17.9 (1.8)  & 49.8 (2.2) \\
& 500 & 17.1 (1.8) & 17.0 (1.8) & 17.0 (1.8) & {\bf 16.6} (1.7) & 29.2 (3.1) & 16.7 (1.7) & 38.6 (13.2) \\
\multirow{2}{*}{(ii)} & 200 & {\bf 2.9} (1.1) & {\bf 2.9} (1.0) & {\bf 2.9} (0.9) & 45.5 (2.8) & 45.7 (2.9) & 10.4 (3.8)  & 20.5 (20.0) \\
& 500 &  {\bf 1.1} (0.5) & {\bf 1.1} (0.5) & 1.2 (0.5) & 44.4 (3.4) & 46.7 (2.5) & 10.0 (5.4) & 13.8 (20.9)\\
\multirow{2}{*}{(iii)} & 200 & 9.7 (1.8) & 9.6 (1.9) & {\bf 9.5} (1.9) & 28.2 (2.6) & 23.4 (3.3) & 20.8 (3.1) & 35.3 (11.5)\\
& 500 & 5.1 (1.3) & 5.1 (1.2) & {\bf 5.0} (1.2) & 26.6 (2.3) & 23.5 (2.7) & 18.7 (2.8) & 33.7 (11.6) \\
\multirow{2}{*}{(iv)} & 200 & 16.0 (2.3) & 16.1 (2.2) & {\bf 15.9} (2.1) & 26.6 (2.5) & 26.5 (2.6) & 26.4 (2.4)  & 27.1 (5.6)\\
& 500 &  13.4 (1.6) & 13.3 (1.8) & {\bf 13.2} (1.7) & 26.5 (2.0) & 26.5 (2.0) & 26.5 (2.0) & 16.5 (5.5)\\
\multirow{2}{*}{(v)} & 200 & 30.9 (2.3) & 31.1 (2.4) & 31.0 (2.4) & 34.9 (2.5) & {\bf 28.3} (2.1) & {\bf 28.3} (2.1) & 38.9 (4.7) \\
& 500 &  27.6 (2.4) & 27.8 (2.3) & 27.8 (2.3) & 33.0 (2.3) & {\bf 23.8} (2.5) & {\bf 23.8} (2.6) & 36.9 (4.1) \\
\hline
\end{tabular}
}
\end{table}

We generate training data following each of models (i) to (v) above with two sampling settings $(n,J)=(200,50)$ and $(500,100)$, while an independent sample of size 500 is generated for testing. The ridged local linear estimator with the plug-in bandwidth as in Section~\ref{sc_est_fd} is applied to obtain smooth curves before applying any method. We set $k_{\rm{PCA}}=15$ and apply the nested ten-fold CV procedure described in~Section~\ref{sc_est_tp} to select $\xi$ and $h_{\rm{reg}}$. The intrinsic dimension $d$ is set to true values for models (i) to (iv) and estimated for model (v) by the minimal neighborhood method \citep{Facco2017} summarized in Section B of the Appendix. For any method involving FPCA, we choose the minimal number of PCs to explain at least 95\% of the variance. We apply all the methods to each of the settings repeatedly for 200 times, and the misclassification errors are summarized in Table~\ref{tb1}. 

We see from Table~\ref{tb1} that the CC performs the best under model (i), closely followed by the NB and FSML methods. The FDNN performs poorly under model (i). Under models (ii) to (iv), the FSML methods significantly outperform the benchmark methods. In particular, the FSML methods achieve near perfect classification under model (ii), where the CC and FQDA perform poorly. It is worth noting that, although we assume nothing about the underlying data distribution, our methods still perform the best under the low-dimensional Gaussian model (iv), while FQDA and NB indeed perform better than ours under the high-dimensional Gaussian model (v). These observations indicate that the FSML methods are particularly efficient if the intrinsic dimension is low, while the FQDA and NB are more powerful when the intrinsic dimension is high, as their asymptotic perfect classification results require the number of PCs tends to infinity. A similar observation has also been noticed in \citet{Wang2024}. We also note that the differences among the three FSML methods are tiny, indicating that the classification performance of the FSML is insensitive to the chosen specific multivariate classifier. Finally, we recall from Remark~\ref{rem2} that the strategy of registration and classification does not perform well under the models above; see Section E of the Appendix for details.

\begin{figure}[t]
\centering
\includegraphics[width=0.32\textwidth]{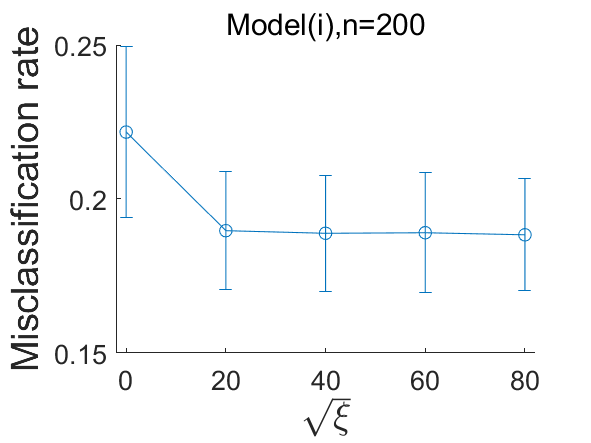}
\includegraphics[width=0.32\textwidth]{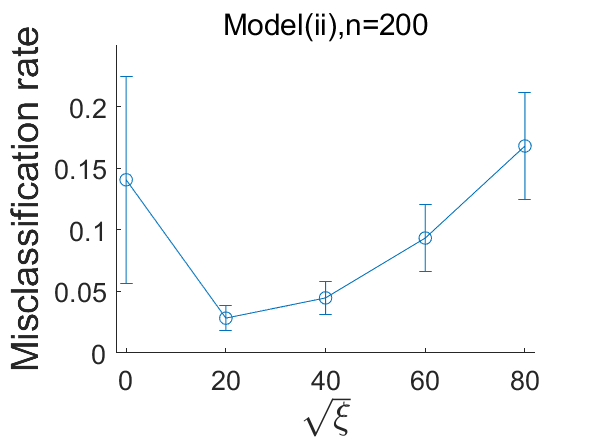}
\includegraphics[width=0.32\textwidth]{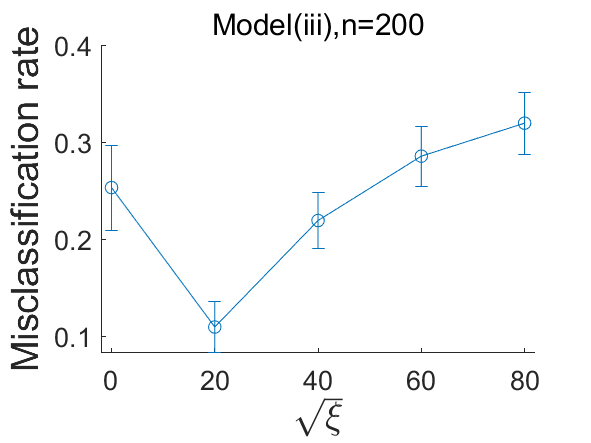}\\
\includegraphics[width=0.32\textwidth]{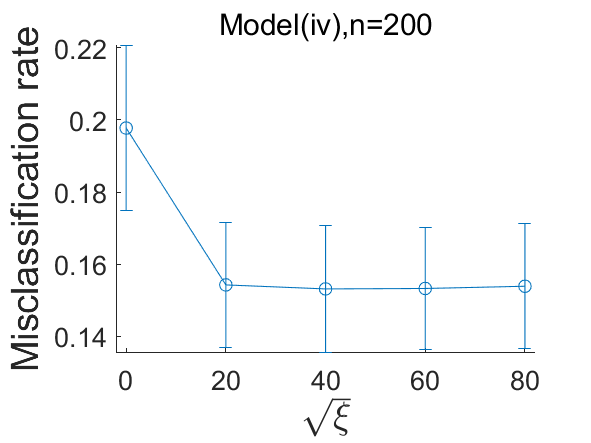}
\includegraphics[width=0.32\textwidth]{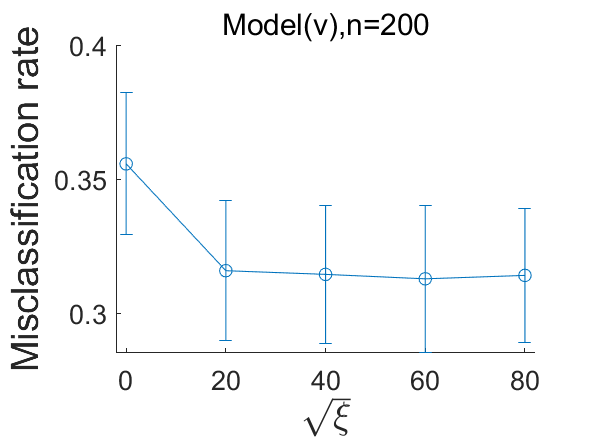}
\caption{The mean misclassification rates ($\circ$) using different values of $\xi$ under all simulation models with $n=200$. The vertical intervals denote the mean $\pm$ standard deviation of the misclassification rates.}\label{fg_diff_xi_1}
\end{figure} 

To see the effect of different values of $\xi$ on the classification performance of our methods, we perform FSML$_{20\rm{NN}}$ on each of the simulation models with different values of $\xi$ and report the results with $n=200$ in Figure~\ref{fg_diff_xi_1}. The corresponding results with $n=500$ are reported in Section E of the Appendix. The parameter $h_{\rm{reg}}$ is chosen via the ten-fold CV and other parameters are set the same as before. Figure~\ref{fg_diff_xi_1} shows that, for models (ii) and (iii), the mean misclassification rates achieve well-separated minima at $\sqrt{\xi} = 20$, while for other models, the mean misclassification rates drop at first but fluctuate little with larger $\xi$. The latter phenomenon is mainly because once the value of $\xi$ is higher than a threshold, the patterns of the manifold learning outcomes stay similar, and thus the classification performance of any classifier does not change much. The reason for presenting $\sqrt{\xi}$ instead of $\xi$ is that the scale of the penalty term in \eqref{eq_def_D} is roughly proportional to $\sqrt{\xi}$. Performances of FSML$_{\rm{SVM}}$ and FSML$_{\rm{LDA}}$ are almost identical to that of FSML$_{k-\rm{NN}}$, and thus they are not reported. 

\begin{figure}[t]
\centering
\includegraphics[width=0.24\textwidth]{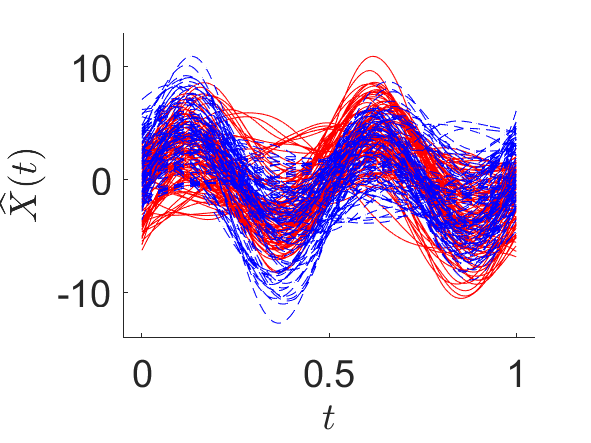}
\includegraphics[width=0.24\textwidth]{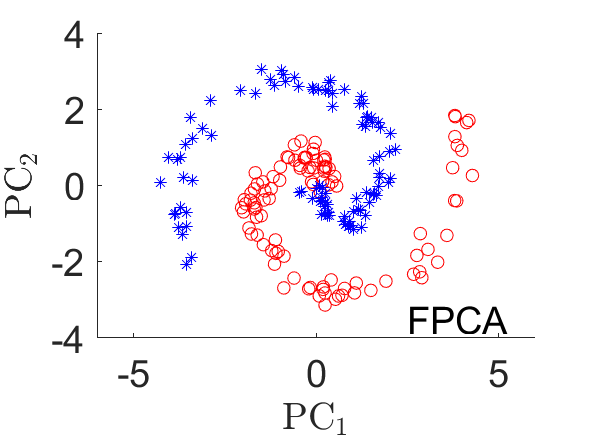}
\includegraphics[width=0.24\textwidth]{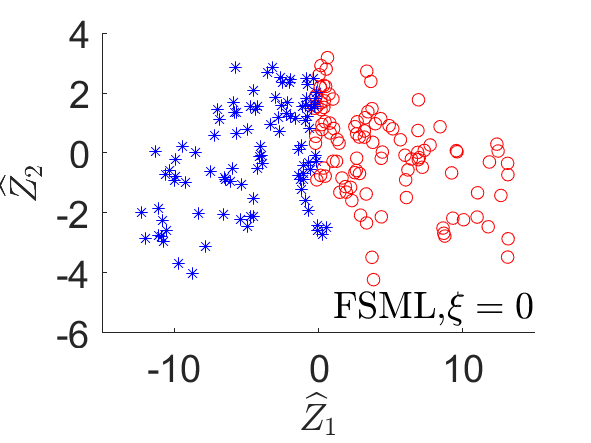}
\includegraphics[width=0.24\textwidth]{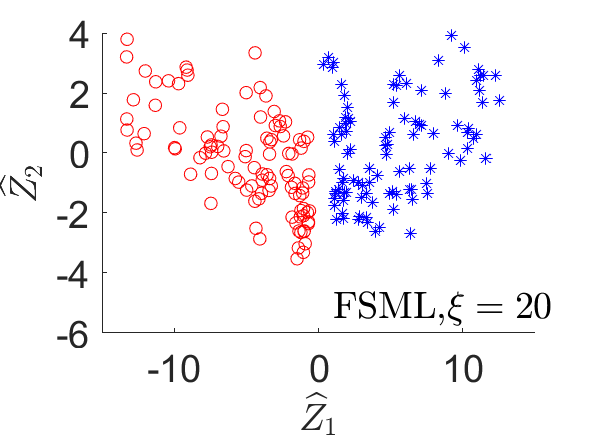}\\
\includegraphics[width=0.24\textwidth]{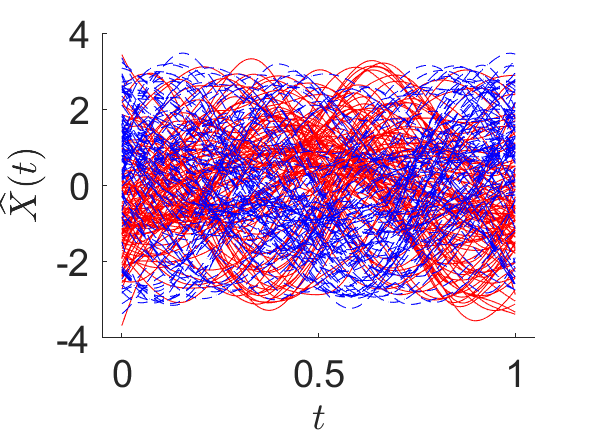}
\includegraphics[width=0.24\textwidth]{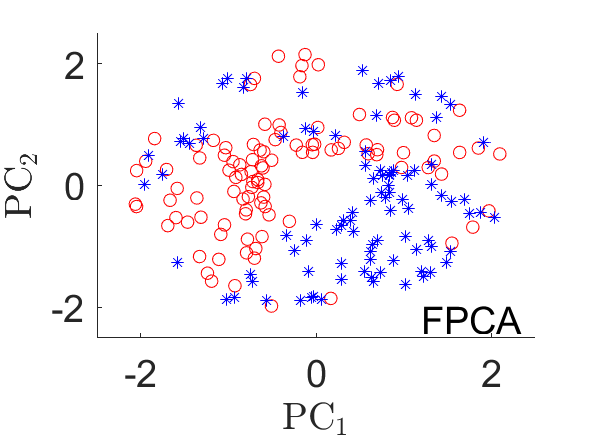}
\includegraphics[width=0.24\textwidth]{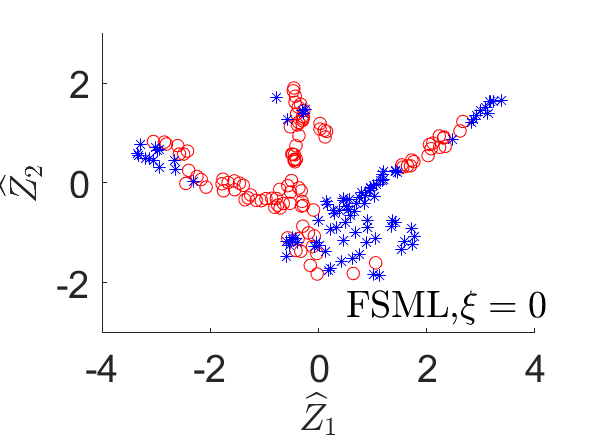}
\includegraphics[width=0.24\textwidth]{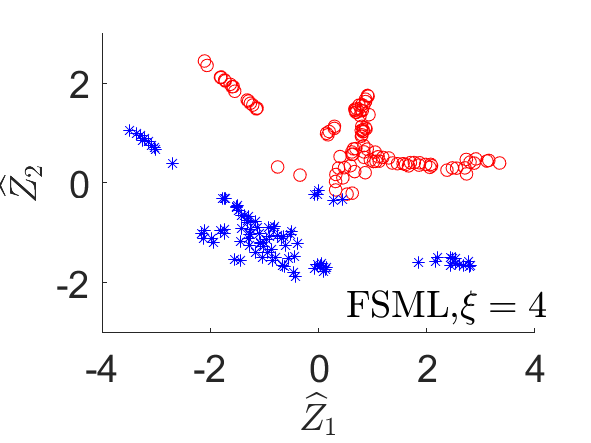}
\caption{The panels from left to right show presmoothed data $\hat{X}_i$, the first two PCs by the FPCA, and manifold learning outcomes using the FSML with different $\xi$'s. The first (resp., second) row is a sample from model (ii) (resp., (iii)) with $n=200$. The individuals with $Y=0$ are denoted by dashed curves and $*$, while those with $Y=1$ are denoted by solid curves and $\circ$.}\label{fg_sim_1}
\end{figure}

One of the applications of the FSML is data visualization. To illustrate this, we generate two samples from models (ii) and (iii), respectively, and apply the FSML, as well as the FPCA for comparison. We show in Figure~\ref{fg_sim_1} the smoothed data by the ridged local linear estimator, the first two PCs by the FPCA, and 2-dimensional manifold learning outcomes by FSML with several $\xi$'s. We see that the FSML properly unfold the Swiss rolls and the outcomes can be linearly separated according to their classes (even with $\xi=0$) under model (ii). Similarly, the FSML outcomes from different classes are also separable with $\xi=4$ under model (iii). In contrast, the first two PCs do not linearly separate two classes under both of the models.

\section{Real Data Applications}\label{sc_real}
We apply all the functional classifiers used in Section~\ref{sc_sim} to three real data examples that are commonly illustrated in the literature on functional data classification.

The first dataset concerns gene expression levels of a yeast cell based on the $\alpha$ factor experiment \citep[][data available at the supplemental material]{Spellman1998}. Several thousand genes were measured every seven minutes from 0 to 119 minutes, of which 612 genes were found to fluctuate periodically within the cell cycle. We focus on the subset that exhibits significant periodic features: the trajectories belonging to the $G_1$ group are coded with $Y=0$ and the trajectories belong to the $G_2$ and $M$ groups are coded with $Y=1$. Here, the groups $G_1,G_2$ and $M$ correspond to different phases of the cell cycle. For example, $G_1$ is known as the first gap phase, during which the cell grows and prepares for DNA replication. We then obtain a sample of 473 trajectories, among which 223 (resp., 250) trajectories belong to $Y=0$ (resp., $Y=1$). Because the observations are quite noisy and the number of observations per gene is moderate, we use the local linear estimator with a manually chosen large bandwidth to recover smooth functional data.

\begin{figure}[t]
\centering
\includegraphics[width=0.3\textwidth]{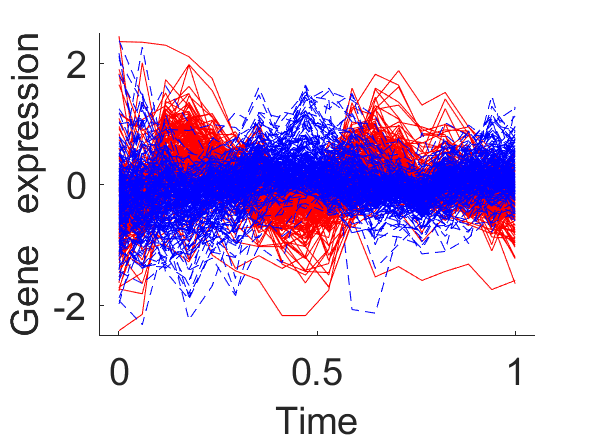}
\includegraphics[width=0.3\textwidth]{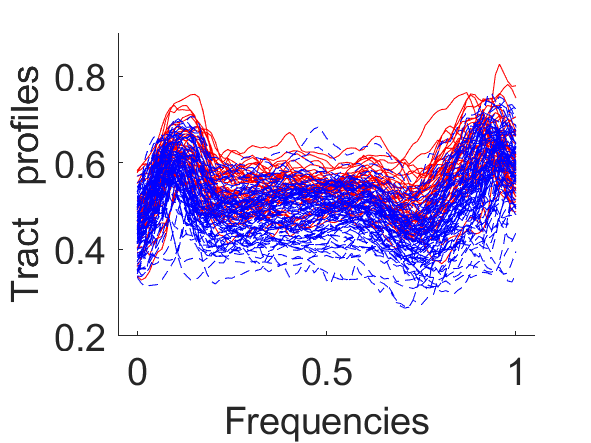}
\includegraphics[width=0.3\textwidth]{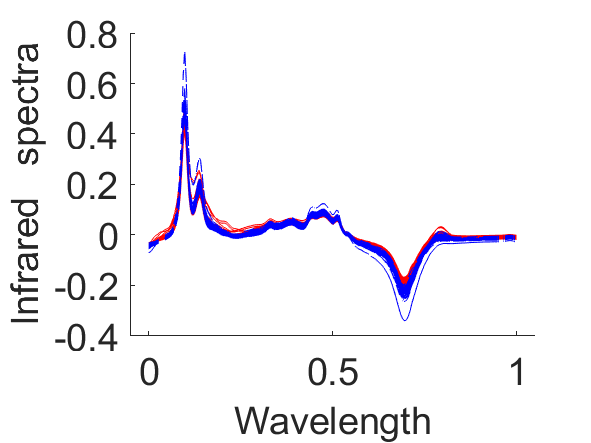} 
\caption{The panels from left to right show raw data of the yeast gene dataset, the DTI dataset and the wine dataset, respectively. The two classes are denoted by solid and dashed lines, respectively. All the domains have been normalized to $[0,1]$. }\label{fg_real_1}
\end{figure}

The second dataset, available at the R package \texttt{refund}, belongs to diffusion tensor imaging (DTI) data collected at Johns Hopkins University and the Kennedy-Krieger Institute \citep{Goldsmith2011}. DTI is an advanced technique commonly used in neuroscience to investigate brain structure and connectivity. The functional trajectories are fractional anisotropy tract profiles from the subject's corpus callosum, which are closely related to multiple sclerosis status. If a subject is measured multiple times, we use the mean tract profiles as one's functional trajectory. The dataset consists of 142 subjects, including 40 healthy cases ($Y=0$) and 102 cases of multiple sclerosis ($Y=1$). The observations are smoothed individually by the ridged local linear estimator with the plug-in bandwidth.

The third dataset consists of wine spectra data, which have been provided by Prof. Marc Meurens, Universit\'e catholique de Louvain (available at the R package \texttt{cggd}). It contains a total number of 123 mean infrared spectra of different wines observed at 256 points. Following \citet{Dai2017}, we regard 45 spectra with alcohol content less than 12 as one group ($Y=0$) and the remaining 78 spectra as the other group ($Y=1$). The data are already smooth, and thus no smoothing is needed.

The raw data for these three datasets are presented in Figure~\ref{fg_real_1}.
To evaluate the classification performance, we perform each of the functional classifiers based on a 5-fold CV on each of the above datasets. The procedure is repeated 100 times, and the mean and standard deviation of the CV misclassification errors in percentage are reported in Table~\ref{tb2}. To implement the FSML methods, the number $k_{\rm{PCA}}$ is set to 10, the intrinsic dimension $d$ is estimated by the minimal neighborhood method \citep{Facco2017}, and $\xi$ and $h_{\rm{reg}}$ are chosen following the same criterion as in Section~\ref{sc_sim}. In addition, graphs of the misclassification rates of FSML$_{5-\rm{NN}}$ with different values of $\xi$ are shown in Section E of the Appendix.

We see from Table~\ref{tb2} that, in the yeast gene dataset, the CC performs the best with a little advantage over the FSML methods, followed by the FQDA and NB, while the FDNN performs less well. Under the DTI dataset, the FSML methods clearly outperform the others. Finally, the FSML methods perform the best under the wine dataset, closely followed by the NB. Overall, the FSML methods demonstrate highly competitive performance compared to other approaches under these three datasets.

\begin{table}[t]
\centering
\caption{Mean (standard deviation) CV misclassification errors in percentage under all real datasets with the best result highlighted in boldface.  }\label{tb2}
\resizebox{\columnwidth}{!}{
%\begin{threeparttable}
\begin{tabular}{cccccccc}
\hline
Dataset  & FSML$_{5-\rm{NN}}$ & FSML$_{\rm{SVM}}$ & FSML$_{\rm{LDA}}$ & CC & FQDA & NB & FDNN\\
\hline
Yeast gene & 5.8 (0.5) & 6.9 (0.7) & 5.7 (0.6) & {\bf 5.5} (0.3) & 7.5 (0.6) & 8.0 (0.6)  & 15.6 (8.3) \\
DTI & {\bf 23.6} (1.6) & 24.1 (1.7) & {\bf 23.6} (1.7) & 26.1 (1.7) & 30.5 (2.1) & 30.2 (2.4)  & 30.0 (2.0) \\
Wine & {\bf 7.0} (1.1) & 7.1 (1.2) & 7.1 (1.5) & 8.7 (1.1) & 10.6 (1.5) & 7.6 (1.2) & 37.6 (2.8)\\
\hline
\end{tabular}
}
\end{table}

\section{Discussion}\label{sc_dis}
In this article, we focus on functional data that lie on an unknown low-dimensional manifold. We propose a novel proximity measure that takes the label information into account, based on which the FSML has been developed. Coupled with multivariate classifiers, the FSML induces a new family of functional classifiers. We provide the convergence rates of the estimated coordinate map and the asymptotic optimality of our functional classifier when the multivariate classifier is the $k$-NN classifier. We show highly competitive classification performance of the FSML methods compared to existing methods across both synthetic and real data examples.

A notable feature of the FSML is that it works particularly efficient when the intrinsic dimension is low, which serves as a motivation for our methodology. In Section~\ref{sc_sim}, we show that the FSML methods outperform several existing methods under two-dimensional manifold models and a Gaussian model of dimension three. Because discrete observations with large noise certainly inflate the estimated intrinsic dimension, presmoothing is important in our procedure.
Note that the intrinsic dimension is equal to the number of PCs when the manifold is flat (i.e., a linear subspace of $L^2(\mathcal{T})$), and thus the low-dimensional intrinsic dimension is, to some extent, in contrast to the infinite number of PCs, the core assumption to show asymptotic perfect classification for functional data \citep{Delaigle2012,Dai2017,Berrendero2018}. In our case, perfect classification is possible if the supports of the distributions $X|Y=y$ are disjoint on the manifold. This situation is not trivial as it seems; see Examples~\ref{exa1} and \ref{exa2} in Section~\ref{sc_model}.

Finally, we remark that it is promising to generalize the FSML to multidimensional and/or multivariate functional data, which requires a proper definition for the intrinsic manifold structure.

\section*{Acknowledgments}
We thank the Editor, the Associate Editor and two referees for their careful reviews and insightful comments that greatly improved our manuscript. Tan's research was supported by the NSFC (No.~12401363 and 12471263), the Fundamental Research Funds for the Central Universities, and the Key Laboratory of Intelligent Computing and Applications (Ministry of Education). Zang's research was supported by the Young Faculty Research Grant (No. 110051360025XN077-56) of North China University of Technology.

\section*{Disclosure Statement}
The authors report there are no competing interests to declare.

\section*{Supplementary Materials}

\begin{description}
    \item[Appendix:] The Appendix ``App\_JCGS-25-043.pdf'' contains technical proofs, further implementation details, and additional experimental results.

    \item[Code and datasets:] The code and datasets can be used to reproduce the numerical results in Sections~\ref{sc_sim} and \ref{sc_real}. Please refer to the ``Readme.pdf'' file to see the instruction of using them. The code is also available at \url{https://github.com/ruoxut/FunctionalManifoldLearning}.
\end{description}

\bibliography{Ref_FSML}

\end{document}